# Detection of large-scale X-ray bubbles in the Milky Way halo

P. Predehl[1]†, R. A. Sunyaev[2,3]†, W. Becker[1,4], H. Brunner[1], R. Burenin[2], A. Bykov[5], A. Cherepashchuk[6], N. Chugai[7], E. Churazov[2,3]†, V. Doroshenko[8], N. Eismont[2], M. Freyberg[1], M. Gilfanov[2,3]†, F. Haberl[1], I. Khabibullin[2,3], R. Krivonos[2], C. Maitra[1], P. Medvedev[2], A. Merloni[1]†, K. Nandra[1]†, V. Nazarov[2], M. Pavlinsky[2], G. Ponti[1,9], J. S. Sanders[1], M. Sasaki[10], S. Sazonov[2], A. W. Strong[1] & J. Wilms[10]

[1]Max-Planck-Institut für Extraterrestrische Physik, Garching, Germany.

[2]Space Research Institute of the Russian Academy of Sciences, Moscow, Russia.

[3]Max-Planck-Institut für Astrophysik, Garching, Germany.

[4]Max-Planck-Institut für Radioastronomie, Bonn, Germany.

[5]Ioffe Institute, St Petersburg, Russia.

[6]M. V. Lomonosov Moscow State University, P. K. Sternberg Astronomical Institute, Moscow, Russia.

[7]Institute of Astronomy, Russian Academy of Sciences, Moscow, Russia.

[8]Institut für Astronomie und Astrophysik, Tübingen, Germany.

[9]INAF-Osservatorio Astronomico di Brera, Merate, Italy.

[10]Dr. Karl-Remeis-Sternwarte Bamberg and ECAP, Universität Erlangen-Nürnberg, Bamberg, Germany.

†e-mail: churazov@iki.rssi.ru; gilfanov@iki.rssi.ru; am@mpe.mpg.de; knandra@mpe.mpg.de; predehl@mpe.mpg.de; sunyaev@iki.rssi.ru

**The halo of the Milky Way provides a laboratory to study the properties of the shocked hot gas that is predicted by models of galaxy formation. There is observational evidence of energy injection into the halo from past activity in the nucleus of the Milky Way[1–4]; however, the origin of this energy (star formation or supermassive-black-hole activity) is uncertain, and the causal connection between nuclear structures and large-scale features has not been established unequivocally. Here we report soft-X-ray-emitting bubbles that extend approximately 14 kiloparsecs above and below the Galactic centre and include a structure in the southern sky analogous to the North Polar Spur. The sharp boundaries of these bubbles trace collisionless and non-radiative shocks, and corroborate the idea that the bubbles are not a remnant of a local supernova[5] but part of a vast Galaxy-scale structure closely related to features seen in γ-rays[6]. Large energy injections from the Galactic centre[7] are the most likely cause of both the γ-ray and X-ray bubbles. The latter have an estimated energy of around $10^{56}$ erg, which is sufficient to perturb the structure, energy content and chemical enrichment of the circumgalactic medium of the Milky Way.**

eROSITA[8] is a large-collecting-area and wide-field-of-view X-ray telescope, launched into space onboard the Spektr-RG mission on 13 July 2019. Over the course of six months (December 2019–June 2020), Spektr-RG and eROSITA have completed a survey of the whole sky at energies of 0.2–8 keV—much deeper than the only other all-sky survey with an X-ray-imaging telescope, which was performed by ROSAT in 1990 at energies of 0.1–2.4 keV.

The sky map from the first eROSITA all-sky survey is shown in Fig. 1. This image has been created from calibrated events in the energy range 0.3–2.3 keV (Methods). A preliminary analysis indicates that more than one million X-ray point sources and about 20,000 extended ones are detected in the survey. This is comparable to, and may exceed, the total number of X-ray sources known before eROSITA launched. Multiwavelength identifications using the WISE and Gaia catalogues[9,10] suggest that about 80% of the point sources are distant active galactic nuclei (AGN; comprising about 80% of all known blazars) and that around 20% are coronally active stars in the Milky Way, including about 150 planet-hosting stars (roughly 10% of all known outside of the Kepler field).

Various very large and diffuse extended structures are visible in the all-sky survey map. The most obvious is a quasi-circular feature, which is part of the North Polar Spur and Loop I (northwest quadrant) discovered in the early days of X-ray and radio astronomy, respectively[5,11].

Although less evident at first glance, close inspection of the medium-energy-band (0.6–1.0 keV) image in the hemisphere below the plane of the Milky Way ('south') reveals an astonishing new feature—a huge circular annulus of similar shape and scale to the structure seen in the north (Fig. 2). Together, they seem to form a pair of 'bubbles' that emerge from the Galactic centre. They are traceable at various levels of intensity throughout most of the sky, and should represent a very large object (several kiloparsecs), akin to the Fermi bubbles, because local features are unlikely to exhibit the fourfold symmetry around the direction towards the centre of the Galaxy.

The Fermi bubbles were discovered in 2010[1] with the Fermi-LAT (Fermi large-area telescope) γ-ray instrument. They have a hard, non-thermal spectrum, which shows up clearly in maps at energies of more than 1 GeV. Their emission is probably due to inverse Compton scattering of cosmic-ray electrons on the cosmic microwave background and other radiation fields. This kiloparsec-scale structure was quickly interpreted as a possible manifestation of past activity of the now dormant supermassive black hole in the centre of the Milky Way, thus linking it with AGN observed outside the Galaxy[12–15]. Alternatively, a burst of star formation could power the bubbles[16–18]. In either case, the energy needed to power their formation must have been very large, at roughly $10^{55}$ erg[1,19].

X-ray emission from the North Polar Spur had already been found by ROSAT[5]. Although considered in most early models to be a nearby supernova remnant nearly surrounding us, the possibility that the North Polar Spur is of Galactic scale has been proposed[6,20], and is supported by several observational arguments[7,21]. In particular, study of absorption in X-ray and radio bands places a lower limit of 300 pc on the distance to the structure[21], which rules out a nearby supernova remnant. In addition, evidence for a large-scale bipolar wind has been presented, based purely on X-ray and mid-infrared data, even before the discovery of the Fermi bubbles[22].

With the eROSITA data, the full scope and morphology of these gigantic X-ray structures has become evident. ROSAT, owing to a combination of its lower sensitivity and softer energy response, could reveal only the brightest part of the southern loop closest to the Galactic plane[1,22], not the whole structure. More recently, the 0.7–1-keV all-sky map from MAXI/Solid-state Slit Camera also provided evidence of a southern enhancement on these large scales, and a close north–south symmetry[23].

The Fermi bubbles and large-scale X-ray emission revealed by eROSITA show remarkable morphological similarity. We therefore suggest that the Fermi bubbles and the eROSITA structure are physically related, and refer to the latter as 'eROSITA bubbles'. Our discovery confirms the previously suggested common origin of the two objects[6,7]. The motivation for a separate name is that, despite the probably common origin, the two structures differ in some important respects.

First, we compare their morphologies on the sky (Fig. 3). The Fermi bubbles are roughly elliptical, about 55° × 45° (north–south, east–west) in diameter, symmetric about the Galactic centre, with vertical axis perpendicular to the Galactic plane, and roughly uniform in γ-ray intensity. The eROSITA bubbles appear as extended as 80° in longitude, roughly 80°–85° in latitude and concentrated in annuli or shells. This suggests that they are, to first-order, close to spherical, with a radius of about 6–7 kpc along the plane, extending radially on the Milky Way close to the Sun, so that their northern and southern edges are imprinted by the closer rim of the bubble. The full vertical extent of the eROSITA bubbles is more difficult to determine; assuming a spherical geometry, they would extend roughly 14 kpc above and below the Galactic plane (Extended Data Fig. 1).

Second, from a preliminary spectral analysis of eROSITA data, the absorbing column density of the diffuse emission in the southwestern bright rim of the eROSITA bubbles (region 'd' in Fig. 2) can be constrained to $N_H = (1.0–3.5) \times 10^{21}$ cm$^{-2}$, consistent with what has been measured previously[21] for the northern structure. One-dimensional cross-sections of the observed surface brightness at various latitudes (Fig. 2) are qualitatively consistent with the projection of (quasi-)spherical thick shells with an outer diameter of 14 kpc. Regardless of the uncertainties on these numbers, it is clear that the eROSITA bubbles are comparable in size to the Galactic disk[24].

We note that the extended X-ray emission revealed by eROSITA coincides spatially with the soft component of the GeV emission reported to surround the Fermi bubbles[2,7,25]. A possible connection with polarized radio-continuum emission at 2.3 GHz and 23 GHz[26] has yet to be explored.

An episodic or continuous energy release in the region of the Galactic centre is expected to generate a series of distinct structures: shocks and contact discontinuities. We see two prominent structures in our maps: one is the outer boundary of the eROSITA bubbles; the other separates the eROSITA bubbles and the Fermi bubbles. The sharp boundary of the eROSITA bubbles, which appears bright in X-rays and indicates hotter gas than what present ahead of it, clearly traces the presence of a non-radiative (or adiabatic) shock (see Methods for an estimate of the gas cooling time). We associate it with a forward shock linked to the onset of large energy release at the Galactic centre. The nature of the boundary between the eROSITA and Fermi bubbles is less clear. It could be another forward shock (in the case of a sequence of energy releases), a reverse shock, a wind-termination shock or a contact discontinuity. The reverse or termination shock models for the Fermi bubbles would imply an additional contact discontinuity somewhere between the Fermi and eROSITA bubbles, which is not apparent in the data. Instead, we consider the simplest scenario in which the eROSITA and Fermi bubbles are causally connected, with the Fermi bubbles driving the expansion of the eROSITA bubbles and both structures being associated with the same (gradual or instantaneous) energy release in the nuclear region of the Milky Way. In this scenario, the outer boundary of the Fermi bubbles plausibly represents a contact discontinuity that separates the shock-heated interstellar medium from the shocked outflow, and the boundary of the eROSITA bubbles is the shock that propagates through the halo gas. The pressure is thus continuous across the interface between the eROSITA and Fermi bubbles and the total thermal energies of the two features simply reflect their volumes (ignoring the effects of stratification, which may be non-negligible). Given that their characteristic sizes differ by a factor of about 2, the total thermal energy of the eROSITA bubbles is almost 10 times larger than that of the Fermi bubbles.

The observed average X-ray surface brightness of $(2–4) \times 10^{-15}$ erg cm$^{-2}$ s$^{-1}$ arcmin$^{-2}$ in the eROSITA bubbles (Methods), which decreases with Galactic latitude, is in broad agreement with such scenario. The observed surface brightness, integrated over the full extent of the eROSITA bubbles, implies a total luminosity of hot X-ray-emitting plasma of $L \approx 1 \times 10^{39}$ erg s$^{-1}$.

To inflate the eROSITA bubbles, an average luminosity of the order of $10^{41}$ erg s$^{-1}$ during the past tens of millions of years would be required, and could arise from either star-forming or AGN activity in the Galactic centre. As discussed above, the arguments in favour of each interpretation in the context of the Fermi bubbles have been debated extensively. In the case of the eROSITA bubbles, the energetics are such that they are at the limit of what the past starburst activity at the centre of the Milky Way could provide. Alternatively, the eROSITA bubbles could be inflated by a period (about 1–2 Myr) of Seyfert-like activity ($L \approx 10^{43}$ erg s$^{-1}$) of the central supermassive black hole (Sgr A*). The long cooling time of the hot plasma is consistent with such a hypothesis.

The structures seen here are reminiscent of similar effects seen in AGN that host rapidly accreting supermassive black holes[1]. These can inject a vast amount of mechanical energy into the ambient gas, as revealed by radio-bright bubbles embedded in the X-ray cocoons[27]. This process, known as AGN feedback, is seen in objects ranging from individual early-type galaxies, such as Centaurus A[28], to massive clusters, such as A426 (Perseus)[29,30], and is thought to have potentially marked effects on the evolution of galaxies. On the other hand, explosions of supernova associated with star formation yield kinetic energy of the order of $10^{51}$ erg per supernova in the ejecta (also known as stellar feedback), which may drive an outflow from the central region of a galaxy[31]. M82 provides a good example of the latter mechanism[32]. The energetics and the most salient features of the observed eROSITA bubbles are such that neither of the two mechanisms could be excluded a priori.

Irrespective of the specific source of energy, our results corroborate the notion that inactive disk galaxies, such as the Milky Way, have hot plasma in their haloes that is highly perturbed by activity in their disks, demonstrating the presence of a feedback mechanism in apparently quiescent galaxies. Galaxies are thought to grow via the slow recondensation of the hot halo plasma, which was shock-heated during the collapse of the dark-matter halo[33]. The cooling time of the hot plasma in the halo is comparable to the Hubble time, so the process of growing a galaxy is assumed to be steady (apart from mergers) and slow. Here we have direct evidence of the re-heating of such plasma, to considerable heights above the Galactic disk.

The detection of these X-ray bubbles was enabled by the combined capabilities of the eROSITA instrument and the Spektr-RG mission profile. More detailed analysis following accurate calibration of the instrument, substantial increases in data quality from the ongoing sky survey and follow-up observations in other parts of the electromagnetic spectrum will reveal further details of the properties of the eROSITA bubbles and the implications for the structure and evolution of galaxies, including the Milky Way.

**Online content** Any methods, additional references, Nature Research reporting summaries, source data, extended data, supplementary information, acknowledgements, peer review information; details of author contributions and competing interests; and statements of data and code availability are available at DOI 10.1038/s41586-020-2979-0.

Received 10 July 2020; accepted 25 September 2020.

**Publisher's note:** Springer Nature remains neutral with regard to jurisdictional claims in published maps and institutional affiliations.

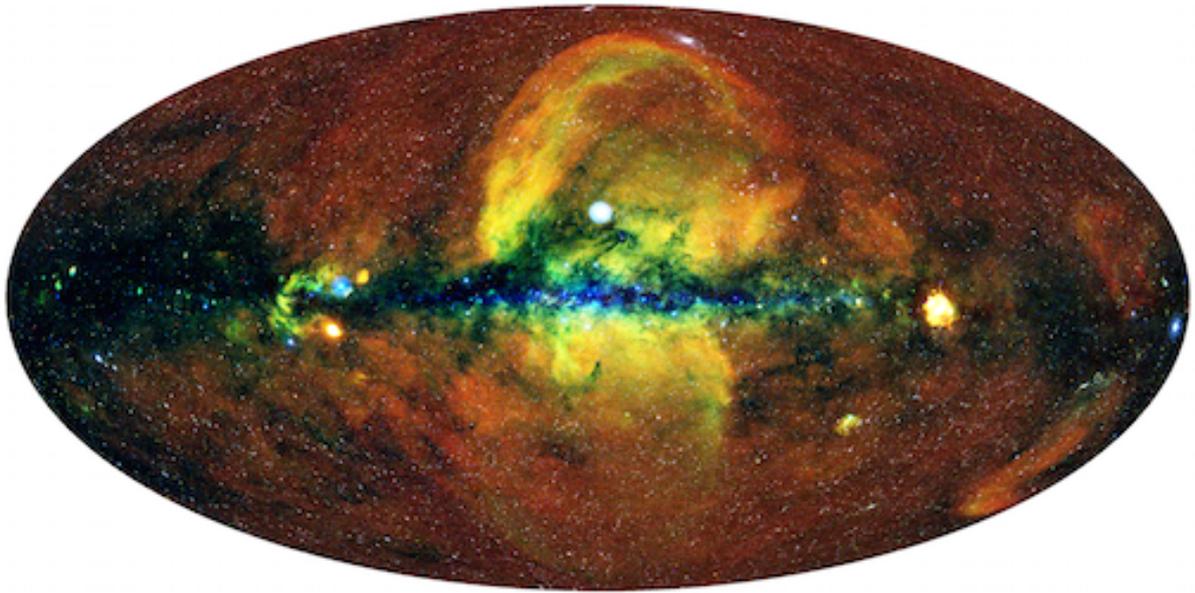

**Fig. 1 | The Spektr-RG–eROSITA all-sky map.** An RGB map of the first Spektr-RG–eROSITA all-sky survey (red for 0.3–0.6 keV, green for 0.6–1.0 keV, blue for 1.0–2.3 keV) is shown in Galactic coordinates, using a Hammer–Aitoff projection. The original image, with a resolution of about 12″, was smoothed (with a Gaussian with a full-width at half-maximum (FWHM) of 10′) to generate this one. CREDIT: Jeremy Sanders, Hermann Brunner, Andrea Merloni and the eSASS team (MPE); Eugene Churazov, Marat Gilfanov (on behalf of IKI).

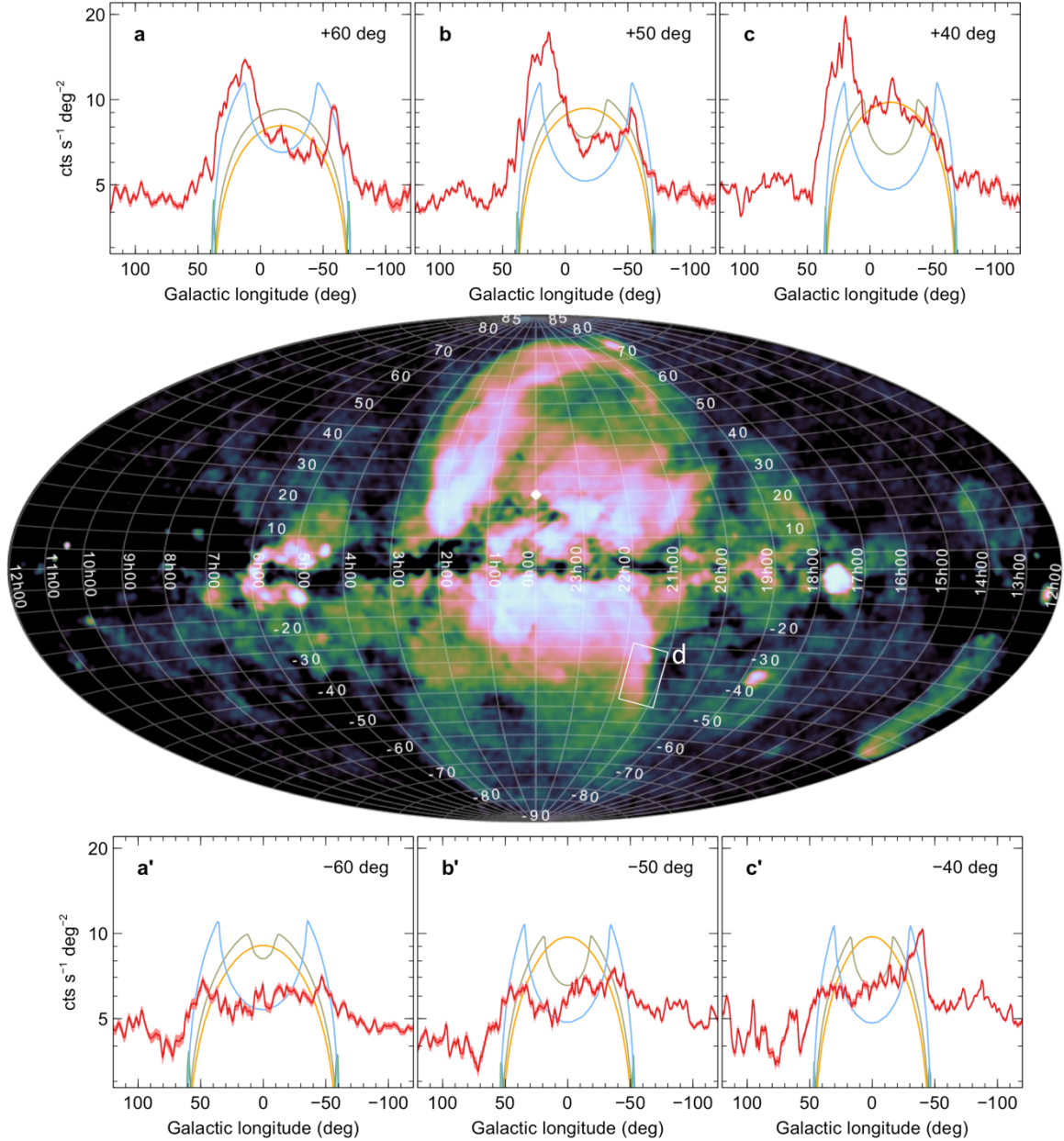

**Fig. 2 | The soft-X-ray eROSITA bubbles. a**, False-colour map of extended emission detected by eROSITA in the 0.6–1.0-keV range. The contribution of the point sources has been removed and the scaling adjusted to enhance large-scale structures in the Galaxy. **b**, One-dimensional surface-brightness profiles in the same energy band (red lines with pink shading representing statistical uncertainties), cut at various galactic latitudes (as labelled). For comparison, we also show the predictions of four possible geometric models (not normalized to the data): a full sphere (yellow), a very thick shell (thickness, 4 kpc; brown), a thick shell (thickness, 2 kpc; cyan) and a thin shell (thickness, 0.2 kpc; green). The thick shell (cyan) is the most consistent with the data (see Extended Data Fig. 2 for a two-dimensional projection of this model). The region indicated by the white rectangle is where a preliminary spectral analysis was performed to constrain the line-of-sight absorption column density towards the southern eROSITA bubbles.

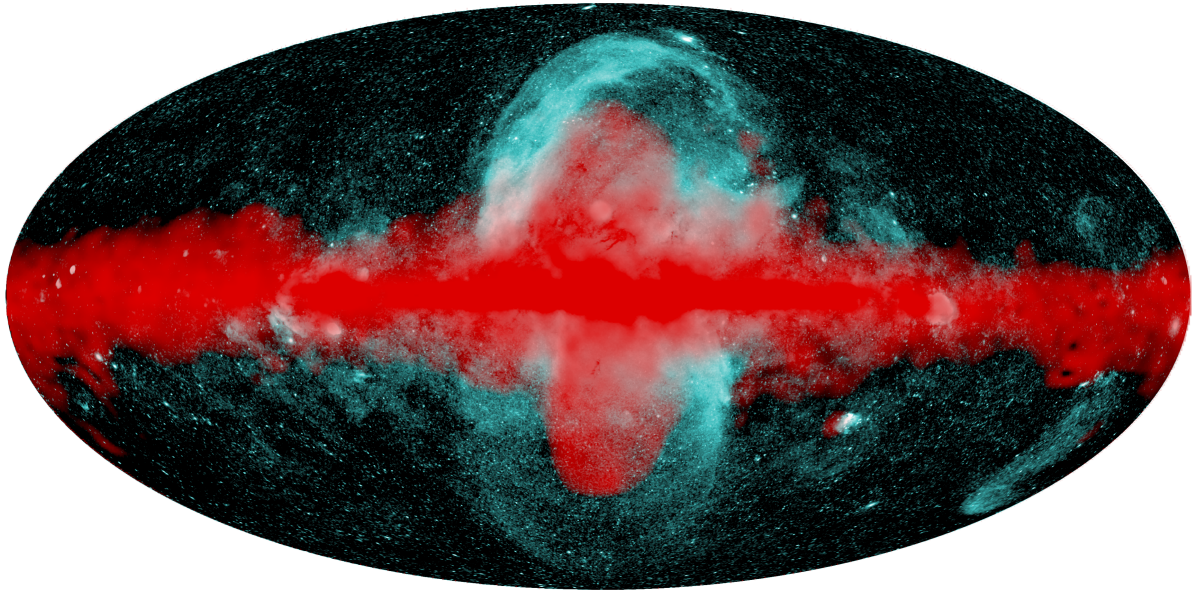

**Fig. 3 | Comparison of the morphology of the γ-ray and X-ray bubbles.** A composite Fermi–eROSITA image is shown. The X-ray extended emission revealed by eROSITA (0.6–1-keV band; cyan) encloses the hard component of the extended gigaelectronvolt emission traditionally referred to as Fermi bubbles (red; Fermi map adapted from ref. [35]), unequivocally establishing their close relation.

## METHODS

### eROSITA aboard Spektr-RG

On 13 July 2019, the X-ray observatory Spektrum-Roentgen-Gamma (SRG; a joint Russian–German mission; R. Sunyaev et al., manuscript in preparation) was launched into space from the Russian cosmodrome Baikonur in Kazakhstan. The Navigator spacecraft platform for SRG (NPO Lavochkin) and launcher (Proton M + Block DM03 upper stage) were both provided by the Russian space agency Roscosmos. The observatory carries two X-ray telescopes: ART-XC (astronomical roentgen telescope X-ray concentrator; M. Pavlinsky et al., manuscript in preparation), a Russian hard-X-ray instrument, and the German-led eROSITA (extended roentgen survey imaging telescope array; P. Predehl et al., manuscript in press). eROSITA was designed to address fundamental questions of astrophysical cosmology such as the nature of dark energy and dark matter, the origin and growth of supermassive black holes, and the expansion of discovery space for rare objects including those of unknown nature[8].

eROSITA consists of seven identical and coaligned X-ray telescopes housed in a common optical bench that connects seven mirror modules with their associated cameras. The dimensions of the telescope structure are approximately 1.9 m diameter and 3.2 m height; the total mass is 810 kg.

The mirror modules comprise 54 paraboloid or hyperboloid mirror shells in a Wolter-I geometry, with an outer diameter of 360 mm and a common focal length of 1,600 mm. The on-axis spatial resolution of all mirror modules is around 16 arcsec (half energy width) at 1.5 keV. X-ray baffles in front of the mirrors effectively suppress stray X-ray light from sources outside the field of view, while magnetic electron deflectors behind the mirrors help to reduce background that arises from low-energy cosmic-ray electrons.

Each mirror module has a p–n charge-coupled device (pnCCD) camera at its focus, with 384 × 384 pixels on a single-chip image area of 28.8 mm × 28.8 mm, which provides a circular field of view of 1.03° diameter. The nominal integration time for all eROSITA cameras is 50 ms. For calibration purposes, each camera has its own filter wheel with a radioactive $^{55}$Fe source and an aluminium or titanium target that provides three spectral lines, at 5.9 keV (Mn Kα), 4.5 keV (Ti Kα) and 1.5 keV (Al Kα). During science operations, the CCDs are passively cooled down to −85° C via heat pipes and radiators. Noteworthy features of the detectors include the low and very stable in-orbit background and excellent spectral response at low energies.

Two star-trackers are mounted on eROSITA to ensure accurate boresight and attitude reconstruction. Thanks to these and the excellent stability of the spacecraft, point-source positional accuracy of about 3″ (1$\sigma$) is achieved. The effective area at 1 keV of all seven telescopes together is similar to that of all three XMM-Newton EPIC cameras; the grasp, defined as the product of field of view and average effective area, is about four times larger. Together with the ability of the spacecraft to continuously scan large areas, this makes eROSITA an extremely fast survey machine, capable of imaging uniformly large swaths of sky. On its three-month cruise to its halo orbit around Lagrangian point $L_2$ of the Sun–Earth system, all SRG systems were put into operation and checked. The instruments were calibrated and their performance verified with a series of scientific observations.

**SRG–eROSITA observations**

The first SRG all-sky survey was conducted between 13 December 2019 and 11 June 2020. Eight all-sky surveys are planned in total, each delivering an average exposure with eROSITA of about 200 s/cos(lat), where lat is the ecliptic latitude; the 1-square-degree area around the ecliptic poles is revisited every four hours, accumulating an exposure of about 30 ks per survey.

During the all-sky survey, the spacecraft rotated continuously with a scan rate of 90° h$^{-1}$, giving a four-hour period per revolution. The rotation axis is oriented to the neighbourhood of the Sun, with an average progression in ecliptic longitude of about 1° day$^{-1}$; it thus completes one all-sky survey in half a year. The scan speed guarantees that the angular resolution is not degraded by smearing of the photons during the 50-ms CCD read-out cycle, and provides sufficient overlap between individual scans to enable source variability analysis and homogeneous survey exposure. Since the start of the survey, the eROSITA cameras have been operating with high efficiency, with more than 96% of the observing time resulting in scientifically useful data. The only substantial loss of efficiency is due to single event upsets in parts of the firmware that control the seven cameras, most probably caused by heavy particles (cosmic rays). Mitigation strategies and procedures have since been put in place by the operations teams. This, combined with the very stable particle background along the large halo $L_2$ orbit of SRG and the flexible mission planning, has resulted in a gap-free coverage for the first all-sky survey.

As SRG scans the sky, for each X-ray photon detected the eROSITA instrument records its position on the detector, energy and time tag. These data, along with detector housekeeping and star-tracker information, are telemetered to ground stations in Russia daily. They are immediately sent from these ground stations, via NPOL Lavochkin and IKI (the Space Research Institute of the Russian Acadmy of Science) to the Max Planck Institute for Extraterrestrial Physics (MPE), where they are checked and converted to FITS format. These files are then cut into 4-h intervals, the duration of a great circle scan. A pipeline based on the eROSITA standard analysis software system (eSASS) determines good time intervals, dead times and corrupted events and frames, masks bad pixels, and applies pattern recognition and energy calibration. Finally, star-tracker and gyro data are used to assign celestial coordinates to each reconstructed X-ray photon. These (time-ordered) photons are then projected into sky tiles of size 3.6° × 3.6°, where images and exposure maps are created for downstream analysis.

Individual sources are detected in the sky tiles using a three-step procedure. First, a local sliding-window detection algorithm identifies enhancements, which are excised from the images. Second, adaptively filtered background maps of the source-free regions are created. Third, the sliding-window detection is repeated, and the identified excesses with respect to the background maps provide an input list of source candidates from which a maximum-likelihood point-spread-function fitting algorithm determines the best-fitting source parameters and detection likelihoods. Data from all seven telescope modules are merged in this approach.

**Generation of all-sky maps**

To produce the sky map shown in Fig. 1, we projected the detected X-ray photons onto the sky, applying smoothing using a two-dimensional Gaussian with FWHM = 10 arcmin, significantly larger than either the native resolution of the map (pixel size of 4″) or the point-spread function of the instrument (FWHM ≈ 12″).

Three sky maps were made in three energy bands (0.3–0.6 keV, 0.6–1.0 keV and 1.0–2.3 keV), which split the X-ray events according to the calibrated and recombined amplitude values from the CCDs. Lower-energy events (recorded down to 0.12 keV) were discarded, owing to the presence of artefacts generated by yet-uncalibrated detector noise; higher energy events (2.3–8 keV) are not shown, owing to the lower sensitivity of eROSITA and the effects of the particle background, which greatly reduce the signal-to-noise ratio in this harder band.

The three images were reprojected into Galactic coordinates using a Hammer–Aitoff projection. This projection maintains equal area on the sky. We similarly computed the total exposure time that each pixel on the sky was observed for during the survey, applying the same Gaussian smoothing. This exposure time takes into account various factors that reduce observed photon rates, such as the number of cameras active at any one time, periods removed by screening software, the effect of vignetting as a source moves away from the optical axis of the telescopes, and bad pixels in the detectors.

Images of the three broadband maps were made using a logarithmic scaling between the upper and lower bounds, chosen to show the majority of the large scale structure. These bounds give dynamical ranges of 13, 35 and 25. The three maps were then combined as red (0.3–0.6 keV), green (0.6–1.0 keV) and blue (1.0–2.3 keV) channels to make an RGB image. Brightness and contrast in each channel was adjusted manually.

The 0.3–0.6-keV band (red), reveals large, soft structures such as the 20°-wide Monogem supernova remnant and the Eridanus superbubble located southwest of the Orion nebula, which cover a large fraction of the X-ray sky. Further towards the direction of the Galactic centre, the Vela supernova remnant, together with the overlapping Puppis A and Vela–Junior remnants, appear as prominent bright and extended X-ray sources. Almost symmetric to Vela with respect to the Galactic centre, the Cygnus superbubble and Cygnus loop are most prominent, with emission connecting Cygnus and distant Draco regions. Between these, just north of the Galactic centre, the brightest persistent X-ray source in the sky (Sco X-1) can be seen; owing to its exceptional brightness, a stray-light halo around the source gives it the appearance of an extended object[34].

Although a hint of extended emission associated with the eROSITA bubbles is directly evident in the all-sky map (Fig. 1), its appearance can be enhanced (Fig. 2). First, it appears most clearly in the 0.6–1.0 keV energy band, where both the effective area of eROSITA and the intensity of the roughly 0.3-keV hot bubbles peak, so we use only this band in the analysis. Further improvement may be achieved by removing the contribution of the point sources from the map. Because point-source analysis in the survey is still ongoing, whereas the eROSITA point-spread-function ground calibration is verified against in-flight data, this task was accomplished using the full sky map obtained as described above, using the Montage, scikit-image and astropy packages. In particular, point sources were identified and masked on small areas of the map reprojected to a set of non-overlapping tiles (to avoid elongation of point sources). This was accomplished using a difference-of-Gaussians filter with a radius corresponding to the mean size of a point source in the image. The filtered image was then thresholded using the levels calculated with the threshold_local function in scikit-image (which allows us to leave all large-scale structures unmasked) and the resulting mask was applied to the observed image. The masked image was convolved using a Gaussian kernel with size of five pixels (astropy.convolve) and reprojected to the initial Hammer–Aitoff projection. Finally, the logarithmic intensity scale and cubehelix colour scale were then applied to the image to allow a large dynamic range of observed intensities to be displayed in print. All local manipulations described above affect only the scales comparable with the point-source size and do not affect the large-scale structures discussed in the paper.

To compare the morphology of the eROSITA and Fermi bubbles (Fig. 3), we used the image of a bubble-like component[35] reprojected into a Hammer–Aitoff projection. The final composite was obtained by applying sequential cyan and red colour scales for the eROSITA and Fermi images, respectively.

**Estimating the energetics of the eROSITA bubbles**

In the 0.6–1.0-keV band, average count rates within the northern and southern eROSITA bubbles of 0.0038 photons $s^{-1}$ arcmin$^{-2}$ and 0.0026 photons $s^{-1}$ arcmin$^{-2}$, respectively, were observed. These count rates translate to fluxes of $3.6 \times 10^{-15}$ erg cm$^{-2}$ s$^{-1}$ arcmin$^{-2}$ and $2.5 \times 10^{-15}$ erg cm$^{-2}$ s$^{-1}$ arcmin$^{-2}$, assuming a plasma with temperature $kT = 0.3$ keV (here $k$ is the Boltzmann constant) and abundances 0.2 times solar[19,21,36,37]. Assuming a projected area of the eROSITA bubbles of 35° × 35° × π for each bubble, the total flux is $5 \times 10^{-8}$ erg cm$^{-2}$ s$^{-1}$ and $3.4 \times 10^{-8}$ erg cm$^{-2}$ s$^{-1}$ for the northern and southern bubbles, respectively. Assuming a distance of 10.6 kpc, we derive luminosities of $6.0 \times 10^{38}$ erg s$^{-1}$ and $4.3 \times 10^{38}$ erg s$^{-1}$.

The total energy content of the eROSITA bubbles may be estimated from their sizes and estimates of the density and temperature of the X-ray-emitting gas. We considered two approaches. First, we assumed a Mach number of the shock of $M \approx 1.5$, as follows from the Rankine–Hugoniot condition for the temperature increase from about 0.2 keV outside the bubbles to around 0.3 keV inside[7,19], and adopted an electron density of the upstream halo gas at a distance of 10 kpc from the centre of roughly $4 \times 10^{-4}$ cm$^{-3}$ (ref. [19]), on the basis of the spectral analysis of numerous XMM-Newton and Suzaku lines of sight. For such a Mach number, the downstream pressure is a factor of about 2.6 larger than the upstream pressure, yielding a total energy of two 7-kpc-radius spheres of roughly $8 \times 10^{55}$ erg. The contribution of the kinetic energy of the expanding gas to the total energy is modest. Second, the density of the gas in the eROSITA bubbles may be estimated from the observed X-ray surface brightness (or X-ray luminosity), assuming a plasma with $kT \approx 0.3$ keV, a metallicity roughly 0.2 times solar[36,37] and size along the line of sight of about 5 kpc (or assuming a shell-like geometry of the bubbles; Fig. 2, Extended Data Fig. 2). Assuming that the X-ray emission from the eROSITA bubbles comes from a shell with inner radius of 3 kpc and outer radius of 7 kpc, we estimate the electron density of the hot emitting plasma to be 0.002 cm$^{-3}$ and the thermal energy within each bubble to be $E_{th} \approx 1.3 \times 10^{56}$ erg.

The velocity of the $M \approx 1.5$ shock in a 0.2-keV gas is approximately 340 km s$^{-1}$. This velocity implies a characteristic expansion time to the present size of around 20 Myr, which translates to an energy-release rate of roughly $(1-3) \times 10^{41}$ erg s$^{-1}$. The cooling time[38] of such tenuous hot gas (density $n = 0.002$ cm$^{-3}$ log($T$) = 6.5) is approximately $1.9 \times 10^8$ yr, much longer than the estimated age of the bubbles. Therefore, once heated, the interior of the bubbles will be visible in X-rays for a long time, even after the energy release has ceased.

## Data availability

The datasets analysed during this study are not yet publicly available. Their proprietary scientific exploitation rights were granted by the project funding agencies (Roscosmos and DLR) to two consortia led by MPE (Germany) and IKI (Russia), respectively. The SRG–eROSITA all-sky survey data will be released publicly after a minimum period of 2 years.

**Acknowledgements** This work is based on data from eROSITA, the primary instrument aboard SRG, a joint Russian–German science mission supported by the Russian Space Agency (Roskosmos), in the interests of the Russian Academy of Sciences, represented by its Space Research Institute (IKI), and the Deutsches Zentrum für Luft- und Raumfahrt (DLR). The SRG spacecraft was built by Lavochkin Association (NPOL) and its subcontractors, and is operated by NPOL with support from IKI and the Max Planck Institute for Extraterrestrial Physics (MPE). The development and construction of the eROSITA X-ray instrument was led by MPE, with contributions from the Dr. Karl Remeis Observatory Bamberg and ECAP (FAU Erlangen-Nuernberg), the University of Hamburg Observatory, the Leibniz Institute for Astrophysics Potsdam (AIP), and the Institute for Astronomy and Astrophysics of the University of Tübingen, with the support of DLR and the Max Planck Society. The Argelander Institute for Astronomy of the University of Bonn and the Ludwig Maximilians Universität Munich also participated in the science preparation for eROSITA. The eROSITA data shown here were processed using the eSASS/NRTA software system developed by the German eROSITA consortium. We thank the entire eROSITA collaboration team, in Germany and Russia, who, over many years, have given fundamental contributions to the development of the mission, the instrument and the science exploitation of the eROSITA data. SRG–eROSITA data processing and calibration and data analyses were performed by a large number of collaboration members in both the German and Russian teams, who also discussed and approved the scientific results presented here. This research made use of Montage. It is funded by the National Science Foundation under grant number ACI-1440620, and was previously funded by NASA's Earth Science Technology Office, Computation Technologies Project, under cooperative agreement number NCC5-626 between NASA and the California Institute of Technology. G.P. acknowledges funding from the European Research Council (ERC) under the European Union's Horizon 2020 research and innovation programme (grant agreement number 865637).

**Author contributions** H.B., M.F., C.M. and J.S.S. developed software to process the eROSITA data and processed the German proprietary data that resulted in the all-sky maps. E.C., M.G., I.K., and P.M. processed the Russian proprietary data. H.B., E.C., M.G., C.M. and J.S.S. performed the analysis that resulted in Fig. 1.

V.D., I.K. and J.S.S. performed the image processing that resulted in Figs. 2, 3 and Extended Data Figs. 1, 2. The majority of the text was written by P.P., W.B., M.F., M.G., E.C., G.P., A.W.S., M.S., H.B. and V.D. V.D. and E.C. worked on Fig. 2; Fig. 3 was created by I.K. and V.D. Extended Data Fig. 1 was prepared by A.M. with the support of an MPE graphic design expert. K.N., A.M. and R.A.S. contributed to writing and editing the manuscript. The above-named authors all contributed to the discussion and interpretation of the results and their implications. The remaining co-authors made important contributions to SRG mission planning and operations, eROSITA data acquisition and analysis, and software development for SRG–eROSITA.

**Competing interests** The authors declare no competing interests.

**Additional information**

**Supplementary information** is available for this paper at https://doi.org/10.1038/s41586-0202979-0.

**Correspondence and requests for materials** should be addressed to P.P., R.A.S., E.C., M.G., A.M. and K.N.

**Peer review information** *Nature* thanks Jun Kataoka and Roland Crocker for their contribution to the peer review of this work. Peer reviewer reports are available.

**Reprints and permissions information** is available at www.nature.com/reprints.

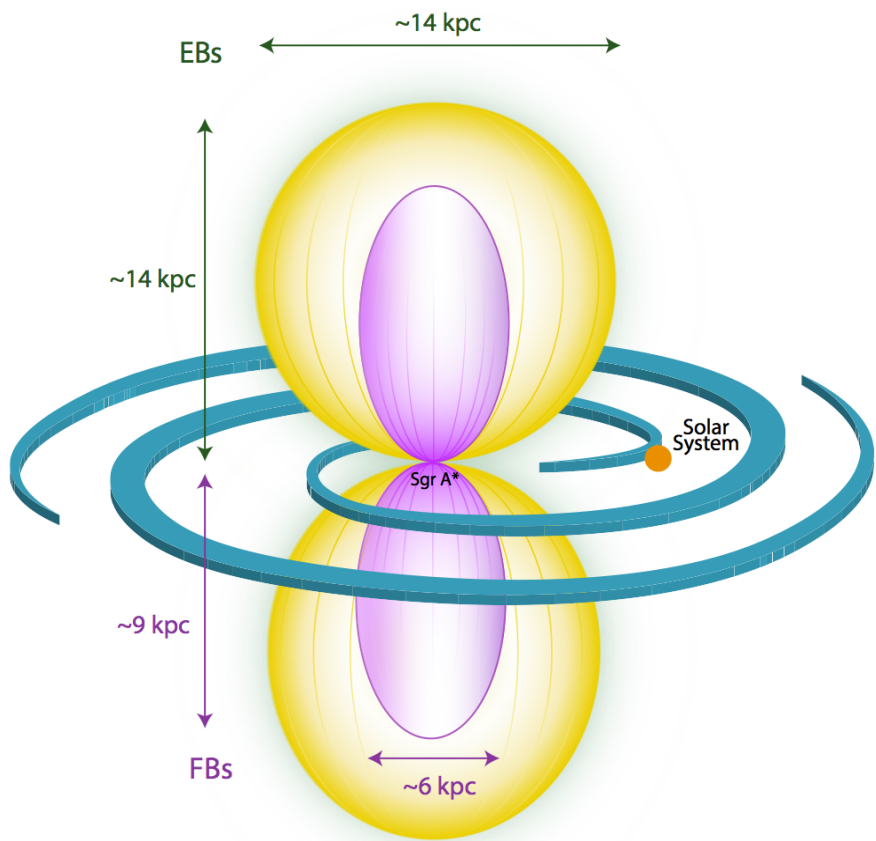

**Extended Data Fig. 1 | Schematic of the eROSITA and Fermi bubbles.** Schematic of the geometry of the eROSITA bubbles (EBs; yellow) and Fermi bubbles (FBs; purple) with respect to the Galaxy and the Solar System. The approximate sizes of these structures, as derived from our analysis, are also marked (green and purple arrows).

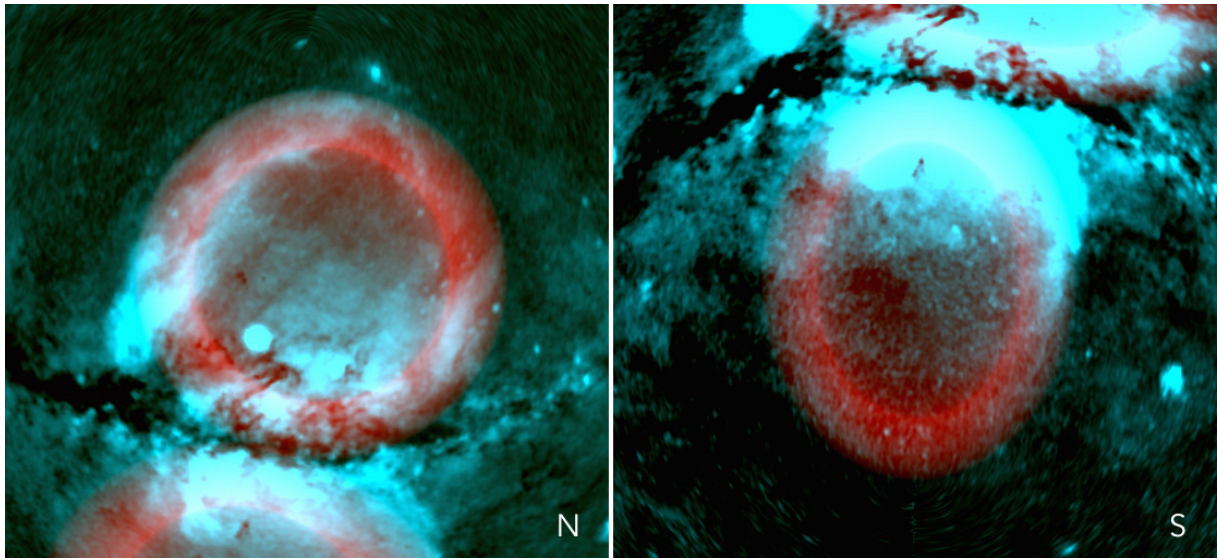

**Extended Data Fig. 2 | Soft-X-ray data compared to a thick-shell model for the eROSITA bubbles.** Comparison between the thick-shell model (cyan line in Fig. 2) and eROSITA data (0.6–1.0-keV band) in a Lambert zenithal equal-area projection. The model is in red; the data are in cyan. The northern bubble is shown on the left (N); the southern bubble is shown on the right (S). The northern bubble is spherical, with an outer radius of 7 kpc and an inner radius of 5 kpc. It is slightly offset from the vertical above the Galactic centre. The southern shell is instead an ellipse, slightly elongated in the north–south direction (semi-major axis is 7 kpc; semi-minor axis 4.9 kpc).